%% file: main.tex
\documentclass{article}
\usepackage{spconf,amsmath,graphicx,hyperref}
\usepackage{multirow,cite,lipsum,enumitem}
\title{FastEnhancer: Speed-Optimized Streaming Neural Speech Enhancement}
\name{
    Sunghwan Ahn \qquad Jinmo Han \qquad Beom Jun Woo \qquad Nam Soo Kim
    \thanks{
        This work was supported by the National Research Foundation of Korea(NRF) grant funded by the Korea government(MSIT) (RS-2025-00554289)
    }
}
\address{
    Department of Electrical and Computer Engineering and INMC\\
    Seoul National University, Seoul, South Korea\\
    \texttt{\{shahn, jmhan, bjwoo\}@hi.snu.ac.kr, nkim@snu.ac.kr}
}

\begin{document}
\ninept
\maketitle
\begin{abstract}

\input{0.Abstract}
\end{abstract}
\section{Introduction}

\input{1.Introduction}

\section{Related Works}
\input{2.Related}

\section{Model Architecture}
\label{sec:model}
\input{3.Model}

\section{Experimental Setup}
\label{sec:setup}
\input{4.Setting}

\input{table-performance}

\input{table-ablation}

\section{Results}
\label{sec:results}
\input{5.Results}

\section{Conclusion}

\input{6.Conclusion}

\clearpage

\let\OLDthebibliography\thebibliography
\renewcommand\thebibliography[1]{
  \OLDthebibliography{#1}
  \setlength{\parskip}{0pt}
  \setlength{\itemsep}{0pt plus 0.3ex}
}
\bibliographystyle{IEEEbib}
\bibliography{refs}

\end{document}

%% file: 0.Abstract.tex
Streaming speech enhancement is a crucial task for real-time applications such as online meetings, smart home appliances, and hearing aids. Deep neural network-based approaches achieve exceptional performance while demanding substantial computational resources. Although recent neural speech enhancement models have succeeded in reducing the number of parameters and multiply-accumulate operations, their sophisticated architectures often introduce significant processing latency on common hardware. In this work, we propose \textit{FastEnhancer}, a streaming neural speech enhancement model designed explicitly to minimize real-world latency. It features a simple encoder-decoder structure with efficient RNNFormer blocks. Evaluations on various objective metrics show that \textit{FastEnhancer} achieves state-of-the-art speech quality and intelligibility while simultaneously demonstrating the fastest processing speed on a single CPU thread. Code and pre-trained weights are publicly available\footnote{https://github.com/aask1357/fastenhancer}.

%% file: 1.Introduction.tex
Monaural speech enhancement aims to estimate a clean speech signal from a single-channel mixture corrupted by noise. Deploying enhancement algorithms in real-time applications like live communication or smart home assistants presents a significant challenge, as it demands not only high-quality output but also fast processing speed on resource-limited devices.

Early traditional approaches like spectral subtraction or Wiener filtering \cite{spectral-subtraction} focused on estimating spectral amplitude of clean speech. However, they rely on noise stationarity assumption and often fail in real-world environments. Other techniques exploited the periodicity of voiced speech \cite{adaptive-comb, adaptive-filt}, but they require accurate pitch information. Furthermore, while many of these traditional techniques improve speech quality, they often fail to enhance speech intelligibility \cite{lim-oppenheim-1979}.
In contrast, deep neural network (DNN)-based models have demonstrated a remarkable ability to handle diverse noise types, significantly surpassing traditional methods \cite{deep-se-overview}. Nonetheless, the high computational demands of these models result in significant processing latency on resource-constrained devices, hindering their deployment on real-world products.

\begin{figure}[htp]
    \centerline{\includegraphics[width=1.0\linewidth]{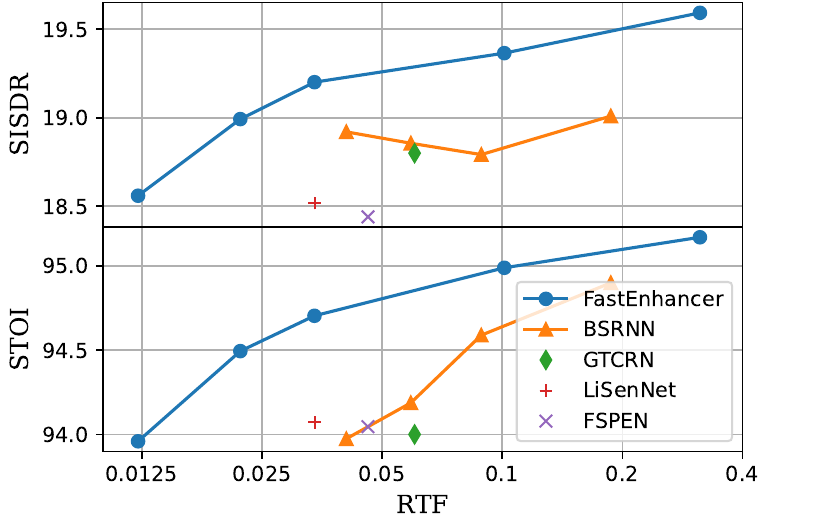}}
    \vspace{-0.4cm}
    \caption{RTF vs. SISDR and RTF vs. STOI of various models.}
    \label{fig:main}
\end{figure}

In response, many recent studies have attempted to decrease the number of parameters or multiply-accumulate (MAC) operations of neural speech enhancement models. These attempts include dual-path recurrent neurat network (DPRNN) structures \cite{dprnn, dpcrn, trunet}, elaborate encoder–decoder designs \cite{gtcrn}, and sub-band decomposition strategies \cite{fspen, lisennet, bsrnn-se, bsrnn-music-separation}. Although these approaches successfully reduce the number of parameters and MAC operations, we observe that they frequently increase memory operations and architectural complexity, which slows down the actual inference speed (Section \ref{sec:related-speed}). Furthermore, despite the widespread adoption of DPRNN, it remains an open question whether it is the optimal structure, particularly concerning its assumption of a sequential relationship between frequency bands (Section \ref{sec:related-dpn}).


To address these limitations, we focus on developing a neural speech enhancement model that satisfies the following properties:
\begin{itemize}[itemsep=0pt,leftmargin=1em]
    \item \textbf{High quality and intelligibility}: The enhanced speech sounds clear and natural (high quality), with the words being easy to understand (high intelligibility).
    \item \textbf{Streaming}: The model can process input spectrograms in a frame-by-frame manner.
    \item \textbf{Fast processing speed}: The model achieves a low real-time factor (RTF) on resource-constrained environments, where RTF is the ratio of processing time to the input audio duration. We consider a single CPU thread as a key benchmark for such environments, as many real-world applications lack dedicated hardware accelerators like GPUs or NPUs.
\end{itemize}

To meet these objectives, we introduce \textit{FastEnhancers}, a family of streaming neural speech enhancement models optimized for speed, quality, and intelligibility. Our key contributions are:
\begin{itemize}[itemsep=0pt,leftmargin=1em]
    \item We demonstrate that a streamlined encoder-decoder structure, in contrast to the complex, sub-band-focused designs of prior works, improves processing speed without sacrificing performance.
    \item We propose an RNNFormer block, an efficient dual-path block that combines a time-axis recurrent neural network (RNN) for low-complexity temporal modeling with a frequency-axis transformer for effectively capturing global, non-sequential relationships across frequency bands.
    \item We evaluate \textit{FastEnhancers} against existing models using various metrics for speech quality and intelligibility on the VCTK-Demand dataset \cite{vctk-demand}. Experimental results show that \textit{FastEnhancers} achieve state-of-the-art performances and the lowest RTFs among all evaluated models. By configuring models with various speeds, we establish a new Pareto frontier for the trade-off between the model performance and inference speed (Fig. \ref{fig:main}).
\end{itemize}

%% file: 2.Related.tex
\subsection{Dual-Path Networks}
\label{sec:related-dpn}
The original DPRNN \cite{dprnn} architecture was proposed for waveform-domain speech separation. It divides an input sequence of $C$ channels and $L$ samples into short chunks with a chunk size of $K$ and a hop size of $P$, resulting in a three-dimensional tensor of shape $[C, L/P, K]$. The model then iteratively performs an intra-chunk RNN along the third dimension and an inter-chunk RNN along the second dimension. This effectively models long sequences and outperforms previous waveform-domain approaches. DPRNN has also been widely adopted for time-frequency (TF)-domain speech enhancement \cite{dpcrn, trunet, fspen, gtcrn, lisennet, bsrnn-se}. It processes a spectrogram by alternating between a time RNN that models temporal dependencies and a frequency RNN that models dependencies across frequency bands. However, applying an RNN to the frequency axis assumes that frequency bands have a sequential relationship. This is suboptimal because spectral relationships (e.g. harmonics) exhibit global rather than sequential dependencies. We argue that a model should be capable of handling all frequency interactions concurrently.

Some prior works have attempted to replace the RNNs in dual-path networks with other powerful modules like Mamba and xLSTM \cite{semamba, xlstm-senet}. However, these models are designed for non-streaming, large-scale scenarios, and still impose a sequential relationship on the frequency axis. Meanwhile, others have explored dual-path transformers (DPT) \cite{dptnet, sepformer}. Transformers are intuitively well-suited for modeling global information across frequency bands. On the other hand, a time-axis transformer is problematic for low-latency streaming. It requires caching previous key and value frames, demanding additional memory operations. Moreover, to improve speech quality, the number of cache frames should be increased \cite{streaming-dpt}, requiring more memory and arithmetic operations.

In this paper, we propose RNNFormer, which is composed of a time RNN and a frequency transformer (Section \ref{sec:model}). The time RNN maintains the low-complexity streaming nature of our model, while the frequency transformer captures global information across all frequency bands. We experimentally prove that RNNFormers achieve a superior RTF-performance curve compared to DPRNNs and DPTs (Section \ref{sec:ablation}).

\subsection{Processing Speed of Speech Enhancement Models}
\label{sec:related-speed}
Many prior works have focused on reducing the number of parameters and MACs to improve model efficiency. DPCRN \cite{dpcrn} and TRU-Net \cite{trunet} introduced a TF-domain U-Net structure with a convolutional encoder, decoder, and DPRNNs. Building upon this, more elaborate methods such as sub-band processing and enhanced DPRNN structures have emerged.

BSRNN \cite{bsrnn-music-separation} introduced a band-split encoder and decoder. FSPEN \cite{fspen} proposed a sub-band encoder and decoder and a time RNN with a path extension. GTCRN \cite{gtcrn} and LiSenNet \cite{lisennet} utilized sub-band feature extraction, grouped DPRNNs, and channel split-concatenate-shuffle operations. While these methods successfully reduced the number of parameters and MACs, their increased architectural complexity and memory operations paradoxically led to higher RTFs (Section \ref{sec:performance}).

In contrast to these approaches, we directly aim to minimize the RTF for practical scenarios. We experimentally prove that a simple network design -- without complex sub-band operations or RNN grouping -- is sufficient to obtain a powerful and fast model.

%% file: 3.Model.tex
\begin{figure*}
    \label{fig:model}
    \centering
    \includegraphics[width=1.0\linewidth]{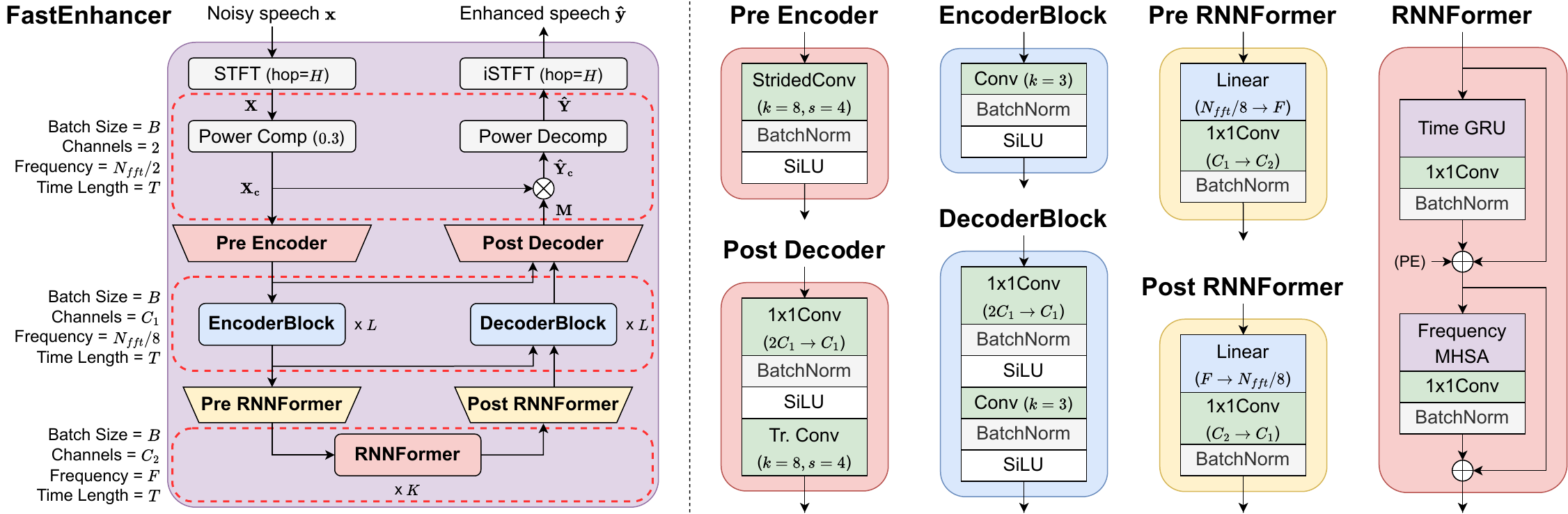}
    \vspace{-20pt}
    \caption{\textit{FastEnhancer} model architecture. Conv($k$), StridedConv($k$, $s$), and Tr. Conv($k$, $s$) denote a convolution, a strided convolution, and a transposed convolution respectively, with a kernel size of $k$ and a stride of $s$. 1x1Conv($A \rightarrow B$) represents a 1x1 convolution with $A$ input channels and $B$ output channels. Linear($A \rightarrow B$) represents a linear layer with $A$ input features and $B$ output features.}
    \vspace{-9pt}
\end{figure*}

This section details the architecture of \textit{FastEnhancer} as shown in Fig. \ref{fig:model}. We start by describing the input and output. We then present convolutional and normalization layers characterized by low latency for streaming inference. Finally, using those building blocks, we explain a simple encoder-decoder structure and our novel RNNFormer blocks. Note that we apply weight normalization \cite{weightnorm} and use the SiLU activation function \cite{silu} across the entire model.

\textbf{Input and Output} -- An input noisy speech $\textbf{x}$ is first transformed into a complex spectrogram $\textbf{X}=\text{STFT}(\textbf{x})$ with a Fourier transform size of $N_{fft}=512$ and a hop size of $H$. To mimic the dynamic compression nature of the human ear, we apply power compression as $\textbf{X}_c = |\textbf{X}|^c\cdot e^{j \angle \textbf{X}}$, where $c=0.3$. We then interpret $\mathbf{X}_c$ as a tensor with two channels (real and imaginary part) and $N_{fft}/2$ frequencies (for simplicity, we discard the highest frequency bin). Given $\mathbf{X}_c$, \textit{FastEnhancer} predicts a two-channel mask \textbf{M} and generates a power-compressed enhanced spectrogram $\hat{\textbf{Y}}_c=\textbf{M} \odot \textbf{X}_c$. After power decompression $\hat{\textbf{Y}} = |\hat{\textbf{Y}}_c|^{1/c}\cdot e^{j \angle \hat{\textbf{Y}}_c}$, the highest frequency bin is zero-padded and the final enhanced speech $\hat{\textbf{y}}=\text{iSTFT}(\hat{\textbf{Y}})$ is synthesized. $\hat{\textbf{y}}$ is then compared to a target clean speech $\textbf{y}$.

\textbf{Convolutional Layers} -- Causal convolutional layers with a time-axis kernel size greater than one introduce latency, because they require caching past frames to compute the current output frame. For a time-axis kernel size of $k$, a buffer of $k-1$ past frames is stored and concatenated with the current frame before the convolution operation. The latest $k-1$ frames are then stored in memory for the next computation. These memory operations significantly increase processing time (Section \ref{sec:ablation}). To ensure minimum latency, \textit{FastEnhancer} exclusively uses convolutions with a time-axis kernel size of one. Throughout this paper, all specified kernel sizes and strides for convolutions refer to the frequency axis.

\textbf{Normalization Layers} -- Speech enhancement models typically use either batch normalization (BN) \cite{batchnorm} or layer normalization (LN) \cite{layernorm}. While both methods introduce negligible number of parameters and MACs, BN can be converted into a simple multiplication-addition operation and fused into adjacent layers at inference time. In contrast, LN cannot be fused in the same manner, increasing the RTF (Section \ref{sec:ablation}). For this reason, we use BN layers throughout our model and fuse them into neighboring layers at inference time.

\textbf{Encoder and Decoder} -- The encoder starts with a pre-encoder network containing a strided convolution that reduces the frequency resolution from $N_{fft}/2$ to $N_{fft}/8$ while enlarging the channels from $2$ to $C_1$. It is followed by a stack of $L$ encoder blocks. The decoder contains $L$ decoder blocks, followed by a post-decoder network that restores the frequency resolution using a transposed convolution. Skip connections are used to link the encoder and decoder.

After the encoder blocks and before the decoder blocks, we append pre-RNNFormer and post-RNNFormer networks to bridge to the RNNFormer blocks. The pre-RNNFormer network is composed of a linear layer that reduces frequency resolution from $N_{fft}/8$ to $F$, followed by a convolution that reduces channels from $C_1$ to $C_2$. The post-RNNFormer network performs the reverse operations. The weight of the linear layer in pre-RNNFormer is initialized as a linear filterbank, and the weight of the linear layer in post-RNNFormer is initialized as a linear interpolation. We empirically found that training these weights offered no performance benefit over keeping them fixed. Therefore, we fixed those weights in all experiments.

\textbf{RNNFormer} -- \textit{FastEnhancer} contains $K$ RNNFormer blocks. Each block consists of two sub-blocks: a time-axis Gated Recurrent Unit (GRU) \cite{gru} block, and a frequency-axis Multi-Head Self-Attention (MHSA) \cite{transformer} block. This hybrid design is critical for low-latency and high-performance. For the time axis, a GRU is well-suited for streaming inference with minimal latency, whereas an MHSA requires a key-value caching that increases delay (as described in Section \ref{sec:related-dpn}). For the frequency axis, an MHSA is ideal as it can model global relationships and doesn't require caching.

The GRU block contains a sequence of a unidirectional GRU, a convolution, a BN layer, and a residual skip connection. The MHSA block follows an identical structure but replaces the GRU layer with an MHSA with four heads. We add a trainable positional encoding to the input of the first MHSA block.

%% file: 4.Setting.tex
We train our model using a composite loss function:
\begin{equation}
    \mathcal{L} = \lambda_{1}\cdot\mathcal{L}_{mag} + \lambda_{2}\cdot\mathcal{L}_{comp} +
                  \lambda_{3}\cdot\mathcal{L}_{con} + \lambda_{4}\cdot\mathcal{L}_{wav} + \lambda_{5}\cdot\mathcal{L}_{pesq},
\end{equation}
where $\lambda_{1}=0.3$, $\lambda_{2}=0.2$, $\lambda_{3}=0.3$, $\lambda_{4}=0.2$, and $\lambda_{5}=0.001$.

\begin{itemize}[leftmargin=1em]
    \item\textbf{Magnitude loss} calculates the mean squared error (MSE) between the magnitudes of $\mathbf{\hat{Y}}_c$ and $\mathbf{Y}_c$:
    \begin{equation}
        \mathcal{L}_{mag} = \left\Vert |\mathbf{Y}_c|-|\mathbf{\hat{Y}}_c| \right\Vert_2^2,
    \end{equation}
    where $\mathbf{Y}=\text{STFT}(\mathbf{y})$ is a spectrogram of a clean speech $\mathbf{y}$ and $\mathbf{Y}_c=|\textbf{Y}|^c\cdot e^{j \angle \textbf{Y}}$ is a compressed spectrogram of $\mathbf{y}$.
    
    \item\textbf{Complex spectrogram loss} interprets two complex spectrograms $\mathbf{\hat{Y}}_c$ and $\mathbf{Y}_c$ as two-channel real-valued tensors, and calculates MSE:
    \begin{equation}
        \mathcal{L}_{comp} = \| \mathbf{Y}_c-\mathbf{\hat{Y}}_c \|_2^2.
    \end{equation}
    
    \item\textbf{Consistency loss} is defined as:
    \begin{equation}
        \mathcal{L}_{con} = \| \mathbf{Y}_c - \text{STFT}_c(\mathbf{\hat{y}}) \|_2^2,
    \end{equation}
    where $\text{STFT}_c(\mathbf{y})= \left| \text{STFT}(\mathbf{y}) \right|^c\cdot e^{j \angle \text{STFT}(\mathbf{y})}$.
    
    \item\textbf{Waveform loss} is a mean absolute error in the waveform domain:
    \begin{equation}
        \mathcal{L}_{wav} = \| \mathbf{y}-\mathbf{\hat{y}} \|_1.
    \end{equation}
    
    \item\textbf{PESQ loss} $\mathcal{L}_{pesq}$ is a differentiable PESQ loss \cite{torch-pesq}. Optimizing a model solely on the PESQ metric can degrade subjective quality \cite{pesqetarian}. To prevent this, we use a sufficiently small weighting factor ($\lambda_5=0.001)$. We observed that including $\mathcal{L}_{pesq}$ improves not only PESQ but also other objective metrics.
\end{itemize}

VTCK-Demand dataset \cite{vctk-demand} was used for training and evaluation, with all audio files downsampled from 48kHz to 16kHz. During training, two-second-long segments were randomly extracted. We used the AdamP optimizer \cite{adamp} with an initial learning rate of 0.002, a weight decay of 0.01, and a batch size of 64. We applied a cosine annealing learning rate scheduler with a 500-step warm-up \cite{coslr}.

\begin{table}
    \vspace{-5pt}
    \caption{Model configurations.}
    \label{table:config}
    \centering
    \begin{tabular}{c|c|c|c|c|c|c}
        \hline
        Size & $H$ & $L$ & $K$ & $C_1$ & $C_2$ & $F$ \\
        \hline
        Tiny (T) & 256 & 2 & 2 & 24 & 20 & 16 \\
        Base (B) & 256 & 2 & 3 & 48 & 36 & 24 \\
        Small (S) & 256 & 3 & 3 & 64 & 48 & 36 \\
        Medium (M) & 160 & 3 & 4 & 96 & 72 & 48 \\
        Large (L) & 100 & 4 & 5 & 128 & 96 & 64 \\
        \hline
    \end{tabular}
\end{table}

We built \textit{FastEnhancers} with five different sizes (Table \ref{table:config}) and compared them with other low-complexity models. We exported every model to ONNXRuntime \cite{onnxruntime} and measured RTFs on a single thread of a server CPU (Intel Xeon Gold 6248R) and a laptop CPU (Apple M1, MacBook Air) under a streaming condition. RTFs in Fig. \ref{fig:main} and Fig. \ref{fig:ablation} are measured on Xeon.

We utilized a variety of objective metrics for evaluation. For speech quality, we measured DNSMOS \cite{dnsmos-p.808} \cite{dnsmos-p.835}, SCOREQ (natural speech, full reference mode) \cite{scoreq}, scale-invariant signal-to-distortion ratio (SISDR) \cite{sisdr}, and PESQ (P.862.2 without Corrigendum 2) \cite{pesq}. For speech intelligibility, we measured STOI \cite{stoi}, ESTOI \cite{estoi}, and word error rate (WER) using Whisper-large-v3-turbo \cite{whisper}. We trained each model five times with five different random seeds and report the average scores.

%% file: table-performance.tex
\begin{table*}[t]
    \caption{Performance on Voicebank-Demand.}
    \label{table:performance}
    \setlength{\tabcolsep}{3.9pt}
    \centering
    \begin{tabular}{c|c|c|c|c|c|c|c|c|c|c|c|c|c|c}
        \hline
        \multirow{2}{*}{Model} & Para. & \multirow{2}{*}{MACs} & RTF & RTF & DNSMOS & \multicolumn{3}{|c|}{DNSMOS (P.835)} & \multirow{2}{*}{SCOREQ} & \multirow{2}{*}{SISDR} & \multirow{2}{*}{PESQ} & \multirow{2}{*}{STOI} & \multirow{2}{*}{ESTOI} & \multirow{2}{*}{WER} \\
        \cline{7-9}
        & (K) & & (Xeon) & (M1) & (P.808) & SIG & BAK & OVL & & & & & & \\
        \hline
        GTCRN$^{\mathrm{a}}$\cite{gtcrn} & \textbf{24} & \textbf{40M} & 0.060 & 0.042 & 3.43 & 3.36 & 4.02 & 3.08 & 0.330 & 18.8 & 2.87 & 0.940 & 0.848 & 3.6 \\
        LiSenNet$^{\mathrm{b}}$\cite{lisennet} & 37 & 56M & - & - & 3.34 & 3.30 & 3.90 & 2.98 & 0.425 & 13.5 & 3.08 & 0.938 & 0.842 & 3.7 \\
        LiSenNet$^{\mathrm{c}}$\cite{lisennet} & 37 & 56M & 0.034 & 0.028 & 3.42 & 3.34 & \textbf{4.03} & 3.07 & 0.335 & 18.5 & 2.98 & 0.941 & 0.851 & 3.4 \\
        FSPEN$^{\mathrm{d}}$\cite{fspen} & 79 & 64M & 0.046 & 0.038 & 3.40 & 3.33 & 4.00 & 3.05 & 0.324 & 18.4 & 3.00 & 0.942 & 0.850 & 3.6 \\
        BSRNN$^{\mathrm{d}}$\cite{bsrnn-se} & 334 & 245M & 0.059 & 0.062 & 3.44 & 3.36 & 4.00 & 3.07 & 0.303 & 18.9 & 3.06 & 0.942 &  0.855 & 3.4 \\
        \textit{FastEnhancer}-B & 92 & 262M & \textbf{0.022} & \textbf{0.026} & \textbf{3.47} & \textbf{3.38} & 4.02 & \textbf{3.10} & \textbf{0.285} & \textbf{19.0} & \textbf{3.13} & \textbf{0.945} & \textbf{0.861} & \textbf{3.2} \\
        \hline
        \textit{FastEnhancer}-T & \textbf{22} & \textbf{55M} & \textbf{0.012} & \textbf{0.013} & 3.42 & 3.34 & 4.01 & 3.06 & 0.334 & 18.6 & 2.99 & 0.940 & 0.850 & 3.6 \\
        \textit{FastEnhancer}-B & 92 & 262M & 0.022 & 0.026 & 3.47 & 3.38 & 4.02 & 3.10 & 0.285 & 19.0 & 3.13 & 0.945 & 0.861 & 3.2 \\
        \textit{FastEnhancer}-S & 195 & 664M & 0.034 & 0.048 & 3.49 & 3.40 & 4.03 & 3.12 & 0.265 & 19.2 & 3.19 & 0.947 & 0.866 & 3.2 \\
        \textit{FastEnhancer}-M & 492 & 2.9G & 0.101 & 0.173 & 3.48 & 3.39 & 4.02 & 3.11 & 0.243 & 19.4 & 3.24 & 0.950 & 0.873 & \textbf{2.8} \\
        \textit{FastEnhancer}-L & 1105 & 11G & 0.313 & 0.632 & \textbf{3.53} & \textbf{3.44} & \textbf{4.04} & \textbf{3.16} & \textbf{0.239} & \textbf{19.6} & \textbf{3.26} & \textbf{0.952} & \textbf{0.877} & 3.1 \\
        \hline
        \multicolumn{15}{l}{$^{\mathrm{a}}$Evaluated using the official checkpoint.} \\
        \multicolumn{15}{l}{$^{\mathrm{b}}$Trained using the official training code. Not streamable because of input normalization and griffin-lim. Thus, RTFs are not reported.} \\
        \multicolumn{15}{l}{$^{\mathrm{c}}$To make the model streamable, input normalization and griffin-lim are removed. Trained following the experimental setup in Section \ref{sec:setup}.} \\
        \multicolumn{15}{l}{$^{\mathrm{d}}$Re-implemented and trained following the experimental setup in Section \ref{sec:setup} for a fair comparison.} \\
    \end{tabular}
\end{table*}

%% file: table-ablation.tex
\begin{table}[htp]
    \vspace{-10pt}
    \caption{Ablation study for convolutions and normalization layers.}
    \label{table:ablation}
    \centering
    \setlength{\tabcolsep}{3.5pt}
    \begin{tabular}{l|c|c|c|c|c}
        \hline
        \multirow{2}{*}{Ablation} & Para. & RTF & RTF & \multirow{2}{*}{SISDR} & \multirow{2}{*}{STOI} \\
         & (K) & (Xeon) & (M1) & & \\
        \hline
        \textit{FastEnhancer}-B & \textbf{92} & \textbf{0.022} & \textbf{0.026} & \textbf{19.0} & \textbf{94.5} \\
        \hline
        \textbf{Convolution} & & & & & \\
        \quad Time-axis kernel size 3 & 187 & 0.028 & 0.037 & \textbf{19.0} & \textbf{94.5} \\
        \hline
        \textbf{Replace BatchNorm to} & & & & \\
        \quad LayerNorm & \textbf{92} & 0.028 & 0.029 & 18.9 & \textbf{94.5} \\
        \hline
    \end{tabular}
\end{table}

%% file: 5.Results.tex
\subsection{Performance Comparisons}
\label{sec:performance}
In Table \ref{table:performance}, \textit{FastEnhancer}-B outperforms all baseline models on almost every metric. While having relatively large number of parameters and MACs, its speed-optimized architecture results in the lowest RTFs. Our smallest model, \textit{FastEnhancer}-T, achieves remarkably low RTFs of 0.012 on Xeon and 0.013 on M1 while maintaining comparable performance to baseline models. As the model size increases, overall performance improves at the cost of increased latency. As a side observation, compared to \textit{FastEnhancer}-S, \textit{FastEnhancer}-M shows slightly degraded DNSMOS scores despite improved scores on every other metric. This indicates that relying on a limited set of objective metrics can lead to misleading conclusions.

\subsection{Ablation Studies}
\label{sec:ablation}
We performed ablation studies to validate the effectiveness of our architectural design. First, we explored the impact of convolution kernel size. Starting with \textit{FastEnhancer}-B, we modified every convolution whose frequency-axis kernel size is 3 in encoder and decoder blocks, enlarging their time-axis kernel size from 1 to 3.
This necessitated caching operations and significantly increased RTFs, but no improvement in quality or intelligibility was observed (Table \ref{table:ablation}).

Second, to investigate the effect of normalization layers, we replaced all BNs with LNs. While this maintained speech quality and intelligibility, it increased RTFs, despite having the same number of parameters and MACs. This result confirms that using BNs and fusing them into adjacent layers is the most efficient option.

\begin{figure}[tp]
    \centerline{\includegraphics[width=1.0\linewidth]{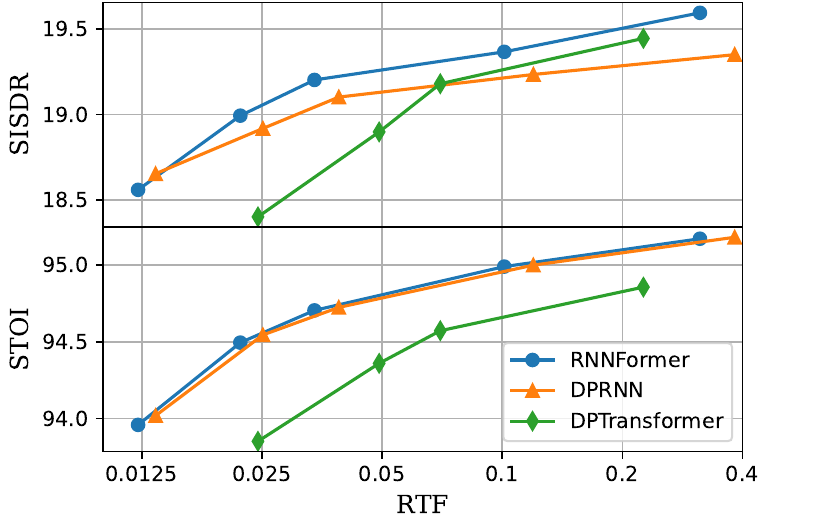}}
    \vspace{-0.4cm}
    \caption{Ablation study for RNNFormer.}
    \label{fig:ablation}
\end{figure}

Third, we created Dual-Path RNNs by replacing the frequency-axis MHSA blocks with frequency-axis GRU blocks.
While STOI scores were similar, DPRNNs clearly degraded SISDR compared to the proposed RNNFormers (Fig. \ref{fig:ablation}).
This demonstrates the superiority of self-attention over a recurrent approach for frequency modeling.

Finally, we constructed Dual-Path Transformers by replacing the time-axis GRU blocks with time-axis MHSA blocks with a look-behind of 31 frames (corresponding to 0.5 seconds). Although DPTs have fewer parameters than RNNFormers, they require key-value caching, resulting in significantly increased RTFs. This result suggests that a recurrent approach is a more efficient choice than self-attention for temporal modeling in a streaming scenario.

%% file: 6.Conclusion.tex
In this paper, we introduced \textit{FastEnhancer}, a streaming neural speech enhancement model designed for low latency. We made careful architectural choices, including convolutional layers with time-axis kernel size of 1, normalization layers that can be fused into other layers at inference, time RNNs, and frequency transformers. Our results demonstrate that \textit{FastEnhancer} significantly outperforms prior works in terms of speech quality and intelligibility while achieving the fastest inference speed on a single CPU thread. This makes our model a highly practical and fast solution for real-world, on-device deployment.